\documentclass[11pt,twoside]{article}
\usepackage{amssymb}
\usepackage{amsmath}
\usepackage{pgf}
\usepackage[latin1]{inputenc}
\usepackage{amsfonts}
\usepackage{amssymb}
\usepackage{amsmath,multirow}
\usepackage{epsfig,array}
\usepackage{graphicx}
\usepackage{url}
\usepackage{hyperref}
\newcommand{\pr} {{\bf Proof. \hspace{0.5cm}}}
\newcommand{\m} {\mathfrak{m}}

\newcommand{\C} {\mathcal{C}}

\newcommand{\F} {\mathbb{F}}

\date{}

\begin{document}

\centerline{}

\centerline{}

\centerline {\Large{\bf Constacyclic codes over} }

\centerline{}

\centerline{\Large{\bf  $\F_q+ u\F_q +v\F_q+ uv \F_q^{~~*}$}}

\centerline{}

\newcommand{\mvec}[1]{\mbox{\bfseries\itshape #1}}

\centerline{\bf {Jo$\ddot{e}$l  Kabor\'e and Mohammed E. Charkani}}

\centerline{}

\centerline{Department of Mathematics, Faculty of Sciences}
\centerline{ Dhar-Mahraz-F$\grave{e}$s, Sidi Mohamed Ben Abdellah University}
\centerline{E-mail: jokabore@yahoo.fr}

\centerline{Department of Mathematics, Faculty of Sciences}
\centerline{Dhar-Mahraz-F$\grave{e}$s, Sidi Mohamed Ben Abdellah University}
\centerline{E-mail: mcharkani@gmail.com}

\centerline{}

\newtheorem{theo}{\quad Theorem}[section]

\newtheorem{definition}[theo]{\quad Definition}

\newtheorem{pro}[theo]{\quad Proposition}

\newtheorem{cor}[theo]{\quad Corollary}

\newtheorem{lem}[theo]{\quad Lemma}

\newtheorem{exa}[theo]{\quad Example}

\centerline{\bf Abstract}
{\emph{Let q be a prime power and $\F_q$ be a finite field. In this paper, we study constacyclic codes over the ring $\F_q+ u\F_q +v \F_q+ uv \F_q$, where $u^2=u, v^2=v$ and $uv=vu.$ We characterize the  generator polynomials of constacyclic codes and their duals using some decomposition of this ring. Finally we study the images of self-dual cyclic codes over $\F_{2^m}+ u\F_{2^m} +v \F_{2^m}+ uv \F_{2^m}$ through a linear Gray map.}}


\section{Introduction}
Constacyclic codes are an important class of linear block codes. These codes possess rich algebraic structures and can be efficiently encoded using shift registers. It's well-Known that for a given unit $\lambda$, $\lambda-$ constacyclic codes over a ring R are ideals of the ring $R[x]/<x^n-\lambda>.$ The last years, these kinds of code have been studied over many classes of finite chain rings \cite{5, 2, 10, 7, 4}. Recently, other classes of rings which are non-chain rings have been introduced. Linear codes and some constacyclic codes over some local frobenius ring have been studied \cite{6,11}. Most recently constacyclic codes over finite principal ideal ring have been investigated \cite{1}. Some results on linear and cyclic codes over the ring $\F_2+u\F_2$ established in \cite{16} have been extended in \cite{3} to the ring $ \F_2[u_1,u_2...,u_k]/<u_i^2-u_i, u_i u_j - u_j u_i>.$ The ring $\F_q+u\F_q+v\F_q+uv\F_q,$ where $u^2=u, v^2=v, uv=vu$ has been used as alphabet to study linear codes and skew-cyclic codes \cite{14}. In the same way, we generalize some results of \cite{15} on constacyclic codes over $\F_q+v\F_q, v^2=v$  to the ring $\F_q+u\F_q+v\F_q+uv\F_q,$ where $u^2=u, v^2=v, uv=vu.$

This paper is organized as follows. In section 2 , we give some properties of the ring $\F_q+u\F_q+v\F_q+uv\F_q,$ and investigate some results about constacyclic codes. In section 3, we characterize the generator polynomials of constacyclic codes, their duals and self-dual constacylic codes over $\F_q+u\F_q+v\F_q+uv\F_q.$ In section 4, we define a gray map over $\F_q+u\F_q+v\F_q+uv\F_q,$ and characterize the Gray images of self-dual cyclic codes.

\section{Preliminaries}
Let R be denote the ring $\F_q+ u\F_q +v\F_q+ uv \F_q$, where $u^2=u, v^2=v$ and $\F_q$ be a finite field with q elements, q is a power of a prime p. This ring is a finite commutative ring with characteristic p and it contains four maximal ideals which are:
$$ \m_1=<u,v>, \m_2=<u-1,v-1>,  \m_3=<u-1, v>, \m_4= <u,v-1> .$$ These ideals have 1 as index of stability. Let
$$\varphi : R \rightarrow R/\m{_1} \times R/\m_2 \times R/\m_3 \times R/\m_4~~ (\cong \F_q^4),$$ be the canonical homomorphism defined by 
$x \longmapsto (x+\m_1, x+\m_2, x+\m_3, x+\m_4).$ By the ring version of the Chinese Remainder Theorem, the map  $\varphi$ is an isomorphism; from this we see that R is a principal ideal ring.\\
We recall a fundamental result on the decomposition of modules.
\begin{lem} [\cite{0}, Proposition 7.2] {~ \newline}
Let R be a finite ring and $I_1, I_2,...,I_n$ be ideals of R. The following statements are equivalent about the R-module R:
\begin{itemize}
\item[i)]   $R= I_1 \oplus I_2 \oplus...\oplus I_n;$
\item[ii)]  There exists a unique family $(e_i)_{i=1}^n$ of idempotents of R such that $e_i e_j=0$ for $i \neq j, 1= \sum_{i=1}^n e_i$ and $I_i =R e_i.$
\end{itemize}
\end{lem} 
Let $e_1 = 1-u-v+uv, e_2 =uv, e_3 = u-uv, e_4= v-uv.$ It is easy to verify that $e_i^2=e_i, e_ie_j=0$ and $1=\sum_{k=1}^{4} e_k$, with $i,j =1,2,3,4,~~ i \neq j$ and $R e_i \cong \F_q$. We deduce, from previous lemma that: $R = Re_1 \oplus Re_2 \oplus Re_3 \oplus Re_4.$
Any element of R can be expressed as: $ r=a+bu+cv+duv = e_1 a +e_2(a+b+c+d)+e_3(a+b)+ e_4(a+c)$, with $a,b,c,d \in \F_q.$
Let :
$$\begin{array}{c c c c}
\varphi :& R &\longrightarrow & \F_q^4 \\
         &r=a+b u+c v+d uv &\longmapsto & (\varphi_1(r), \varphi_2(r), \varphi_3(r),\varphi_4(r))
\end{array}$$
Where
$$\begin{array}{c c c c }
\varphi_1 :& R &\longrightarrow & R/\m_1 \cong \F_q \\
         &r=a+b u+c v+d uv &\longmapsto& a.
\end{array}$$
$$\begin{array}{c c c c }
\varphi_2 :& R &\longrightarrow & R/\m_2 \cong \F_q \\
         &r=a+b u+c v+d uv &\longmapsto & a+b+c+d.
\end{array}$$
$$\begin{array}{c c c c}
\varphi_3 :& R &\longrightarrow & R/\m_3 \cong \F_q \\
         &r=a+b u+c v+d uv &\longmapsto& a+b.
\end{array}$$
$$\begin{array}{c c c c }
\varphi_4 :& R &\longrightarrow & R/\m_4 \cong \F_q \\
         &r=a+b u+c v+d uv &\longmapsto& a+c.
\end{array}$$
By the module version of chinese remainder theorem, $\varphi$ is an $R-$module isomorphism. This map can be extended to $R^n.$
For a code $\C \subseteq R^n,$ we denote $\varphi_i(\C)$ by $\C_i$ for $1 \leq i \leq 4;$ then we have $\C \cong \C_1 \times \C_2 \times \C_3 \times \C_4$ and $|\C| =|\C_1||\C_2||\C_3||\C_4|.$ Note that an element $\lambda = a+bu+cv+duv \in R^*$ is a unit if and only if $\forall i \in \{1,2,3,4\}, \varphi_i( \lambda)$ is a unit in $\F_q$ if and only if $ a \neq 0, a+b+c+d \neq 0, a+b \neq 0$ and $a+c \neq 0.$

The following result is a consequence of theorem 4.9 of \cite{1}.
\begin{lem} \label{i1}
Let $\lambda= a+ bu +cv +d uv$ be a unit in R and $\C=\varphi^{-1}(\C_1 \times \C_2 \times \C_3 \times \C_4)$ be a code of length n over R. Then $\C$ is a $\lambda-$ constacyclic code over R if and only if $\C_1,\C_2,\C_3,\C_4$ are  $a-$ constacyclic, $(a+b+c+d)-$ constacyclic, $(a+b)-$ constacyclic, and $(a+c)-$constacyclic codes of length n over $\F_q$, respectively.
\end{lem}
%
\section{Constacyclic codes over \texorpdfstring{$\F_q + u\F_q +v\F_q+ uv \F_q^{~~*}$}{Fq +uFq+vFq+uvFq}}
Now we investigate constacyclic codes over R. From the previous section, we know that any code over R can be uniquely expressed as $ \C= e_1 \C_1 \oplus e_2 \C_2 \oplus e_3 \C_3 \oplus e_4 \C_4.$ Let $\lambda=a+bu+cv+duv$ be a unit in R. We let  $\lambda_1=a, \lambda_2=a+b+c+d, \lambda_3=a+b$ and $\lambda_4=a+c$.
\begin{theo} \label{i3}
Let $ \C= e_1 \C_1 \oplus e_2\C_2 \oplus e_3 \C_3 \oplus e_4 \C_4$ be a $\lambda-$ constacyclic code of length n over R. Then $ \C= < e_1 g_1(x), e_2 g_2(x), e_3 g_3(x), e_4g_4(x)>,$ where $g_i(x)$ is a generator polynomial of $\lambda_i-$ constacyclic code $\C_i, 1 \leq i \leq 4.$
Furthermore $|\C|= q^{4n- \sum_{i=1}^4 \deg g_i(x)}.$
\end{theo} 
\pr
If $ \C= e_1 \C_1 \oplus e_2\C_2 \oplus e_3 \C_3 \oplus e_4 \C_4$ is a $\lambda-$ constacyclic code of length n over R, then from lemma \ref{i1},$~ \C_i$ is a $\lambda_i-$ constacyclic code of length n over $\F_q.$ So there exists polynomials $g_1(x), g_2(x), g_3(x), g_4(x)$ such that $\C_i=< g_i(x)>,$ for $1 \leq i \leq 4.$\\ 
Let $r(x) \in \C,$ since $ \C= e_1 \C_1 \oplus e_2\C_2 \oplus e_3 \C_3 \oplus e_4 \C_4$, then there exists $f_i(x) \in \C_i=<g_i(x)>, 1\leq i\leq 4 $ such that $r(x)= \sum_{i=1}^4 e_i f_i(x), i.e.$ there exists $h_i(x) \in \F_q[x]$ such that $r(x)=\sum_{i=1}^4 e_i h_i(x) g_i(x).$
Hence $r(x) \in$ $ < e_1 g_1(x), e_2 g_2(x), e_3 g_3(x), e_4g_4(x)>$ i.e. $\C \subseteq < e_1 g_1(x), e_2 g_2(x), e_3 g_3(x), e_4g_4(x)>.$\\
Reciprocally if $r(x) \in < e_1 g_1(x), e_2 g_2(x), e_3 g_3(x), e_4g_4(x)>,$ there are polynomials $k_i(x) \in R[x]/<x^n- \lambda>$ such that $r(x)= \sum_{i=1}^4 e_i g_i(x) k_i(x);$ then there are $r_i(x) \in \F_q[x]$ such that $r(x)= \sum_{i=1}^4 e_i g_i(x)r_i(x)$ where $g_i(x) r_i(x) \in \C_i$  $\subseteq \F_q[x]/<x^n-\lambda_i>;$ therefore $r(x) \in \C $ and $< e_1 g_1(x), e_2 g_2(x), e_3 g_3(x), e_4g_4(x)>\subseteq \C;$ which implies that \\ $\C=< e_1 g_1(x), e_2 g_2(x), e_3 g_3(x), e_4g_4(x)>$. \\
Since $|\C|=|\C_1||\C_2||\C_3||\C_4|$, we deduce that $|\C|= q^{4n- \sum_{i=1}^4 \deg g_i(x)}.$  \hfill $\square$
\\
For any code of length n over R and any r $\in R$, we denote by $(\C :r)$ the submodule quotient defined as follows:
$$(\C:r)=\{s \in R^n | rs \in C\}.$$
%
\begin{lem} \label{i2}
Let $ \C= e_1 \C_1 \oplus e_2\C_2 \oplus e_3 \C_3 \oplus e_4 \C_4$ be a linear code over R. Then:
$$ \varphi_i((\C:e_i))=\C_i, \forall~ 1\leq i \leq 4.$$
\end{lem}
\pr
Let $r \in (\C:e_i),$ then $e_i r \in \C.$  
We can write r as: $r= e_1 \varphi_1(r) + e_2 \varphi_2(r)+e_3 \varphi_3(r)+ e_4 \varphi_4(r);$ then $e_i r =e_i \varphi_i(r), 1 \leq i \leq 4$, which implies that $\varphi_i(r) \in \C_i, 1 \leq i \leq 4$ hence $\varphi_i((\C:e_i)) \subseteq \C_i, 1 \leq i \leq 4.$\\
Reciprocally, for any $r_1 \in \C_1$ there exists $r_2, r_3, r_4 \in \C_2,\C_3, \C_4,$ respectively such that: $e_1 r_1 + e_2 r_2 + e_3 r_3 + e_4 r_4 \in \C$. We see that $e_1 r_1 = e_1(e_1 r_1 + e_2 r_2 + e_3 r_3 + e_4 r_4) \in e_1 \C \subseteq \ $ and $r_1= e_1 r_1+e_2 r_1+e_3r_1+e_4 r_1;$ so $r_1 \in (\C:e_1)$ and $\varphi_1(r_1)=r_1.$ Then $r_1 \in \varphi_1((\C: e_1)),$ hence $\C_1 \subseteq \varphi((\C:e_1))$.
The proof is the same for the other cases.  \hfill $\square$
\\
The following result is a generalization of theorem $3.5$ in \cite{15}.
\begin{theo}
Let $ \C= e_1 \C_1 \oplus e_2\C_2 \oplus e_3 \C_3 \oplus e_4 \C_4$ be a  $\lambda -$ constacyclic code of length n over R. We suppose that 
$ \C= < e_1 g_1(x), e_2 g_2(x), e_3 g_3(x), e_4g_4(x)>,$ where polynomials $g_i(x), 1 \leq i \leq 4$ are monic with $g_i(x)$ divides $(x^n- \lambda_i)$ for $1 \leq i \leq 4.$ Then, for $1 \leq i \leq 4, g_i(x)$ is the generator polynomial of $\lambda_i-$ constacyclic code.
\end{theo}
\pr
For a polynomial $f(x) \in (\C:e_i)$, we have $ e_i f(x) \in \C .$ Since $\C= < e_1 g_1(x), e_2 g_2(x), e_3 g_3(x), e_4g_4(x)>$, then for $1  \leq j \leq 4,$ there exists $s_j(x) \in R[x]/<x^n-\lambda>$ such that $e_i f(x)= \sum_{j=1}^4 e_j g_j(x)s_j(x).$ Furthermore $f(x)=\sum_{i=1}^4 e_i \varphi_i(f(x))$ and $s_j(x)= \sum_{j=1}^4 e_j \varphi_j(s_j(x));$ hence $$e_i[\sum_{j=1}^4e_j \varphi_j(f(x))]=\sum_{j=1}^4 e_j g_j(x)[\sum_{j=1}^4 e_j\varphi_j(s_j(x))].$$
This implies that $e_i \varphi_i(f(x))= \sum_{j=1}^4 e_j g_j(x) \varphi_j(s_j(x)).$ We deduce: $\varphi_i(f(x))=g_i(x)\varphi_i(s_i(x)).$ So 
$\varphi_i(f(x)) \subseteq <g_i(x)>$. Reciprocally, if $f(x) \in <g_i(x)>,$ then there exists $r(x) \in \F_q$ such that $f(x)=g_i(x)r(x).$ This implies that   $e_i f(x)=e_i g_i(x)r(x) \in \C,$ i.e. $f(x) \in (\C: e_i).$ Since $f(x) =\sum_{i=1}^4 e_i f(x),$ then $\varphi_i(f(x))=f(x),$ hence $f(x) \in \varphi_i((\C:e_i)).$
This implies that $<g_i(x)> \subseteq \varphi_i((\C:e_i)).$ The result follows from lemma \ref{i2}.  \hfill $\square$
\\
\begin{theo} \label{i4}
Let $\C$ be a $\lambda-$ constacyclic code over R and $g_i(x)$ be the monic generator polynomial of the code $\C_i, 1 \leq i \leq 4.$ Then there exists a unique polynomial $g(x) \in R[x]$ such that $\C=<g(x)>$ and $g(x)$ is a divisor of $x^n- \lambda.$
\end{theo}
\pr

$\circ$ Let $g(x)=\sum_{i=1}^4 e_i g_i(x).$ It's obvious that  $<g(x)> \subseteq \C.$ Reciprocally, it's clear that $e_i g_i(x) = e_i (\sum_{j=1}^4 e_j g_j(x))= e_i g(x), \forall~ 1 \leq i \leq 4,$ which implies that $\C \subseteq <g(x)>,$ hence $\C= <g(x)>.$

$\circ$ Unicity of $g(x):$ we suppose that there exists another polynomial $h(x)$ in $R[x]/<x^n- \lambda>$ such that $\C=<h(x)>.$ For $1 \leq i \leq 4$, we have $e_i h(x)= e_i[\sum_{j=1}^4 e_j \varphi_j(h(x))]= e_i \varphi_i(h(x)) \in \C.$ This implies $\varphi_i(h(x)) \in \varphi_i((\C: e_i)) = \C_i$ ( from lemma \ref{i2}). Since $\C_i =<g_i(x)>$ , we deduce that $g_i(x)$ divides $\varphi_i(h(x)), 1 \leq i \leq 4.$ Conversely , since $<g(x)>=<h(x)>,$ there exists polynomial $k(x) \in R[x]/<x^n- \lambda>$ such that: 
$$\sum_{i=1}^4 e_i g_i(x)= k(x) h(x)=[ \sum_{j=1}^4 e_j \varphi_j(k(x))][\sum_{j=1}^4 e_j \varphi_j(h(x))]=\sum_{j=1}^4 e_j \varphi_j(k(x))\varphi_j(h(x)).$$ It follows that  
$g_i(x)=\varphi_i(k(x))\varphi_i(h(x))$ i.e. $\varphi_i(h(x))$ divides $g_i(x), \forall ~ 1 \leq i \leq 4.$ Then we conclude that $\varphi_i(h(x))=g_i(x), \forall ~ 1 \leq i \leq 4;$ so $h(x)=g(x)$ in $R[x]/<x^n- \lambda>.$

$\circ$ Now we show that the polynomial $g(x) \in R[x]/<x^n- \lambda>$ is a divisor of $ x^n- \lambda.$ We know that $g(x)= \sum_{i=1}^4 e_i g_i(x)$, where $g_i(x)$ is monic generator polynomial of $\lambda_i-$ constacyclic code $\C_i, 1 \leq i \leq 4.$ Then $g_i(x)$ is a divisor of $x^n- \lambda_i$ in $\F_q[x].$ This implies that there exists $h_i(x) \in \F_q[x]$ such that $x^n-\lambda_i= g_i(x)h_i(x).$ Thus $(\sum_{j=1}^4 e_j g_j(x))(\sum_{j=1}^4 e_j h_j(x))= \sum_{j=1}^4 e_j g_j(x) h_j(x)=\sum_{j=1}^4 e_j(x^n- \lambda_j)=\sum_{j=1}^4 e_j \varphi_j(x^n- \lambda)= x^n- \lambda.$ Therefore $g(x)$ divides $x^n- \lambda.$  \hfill $\square$
\\
As a consequence of previous theorem we have:
\begin{cor}
Let $\lambda$ be an unit in R;
$\frac{R[x]}{<x^n-\lambda>}$ is a principal ideal ring.
\end{cor}
Now we discuss about dual of constacyclic codes over R. Given codewords $r=(r_0,r_1,...,r_{n-1}),$ $s=(s_0,s_1,...,s_{n-1}) \in R^n$, their inner product is defined in the usual way: $$ r.s= r_0s_0+r_1s_1+...r_{n-1}s_{n-1},\text{evaluated in R}.$$
The dual code $\mathcal{C}^\bot$ of $\mathcal{C}$ is the set of n-tuples over R that are orthogonal to all codewords of $\mathcal{C}$, i.e.:
$$\mathcal{C}^\bot=\{r | r.s=0,\forall s \in \mathcal{C}\}.$$
The code $\mathcal{C}$ is called self-dual if $\mathcal{C}=\mathcal{C}^\bot$.\\
It's well-known that for linear codes of length n over a finite frobenius ring R, $|\mathcal{C}||\mathcal{C}^\bot|=|R|^n$ (\cite{13}).
For a given unit $\lambda \in R,$  the dual of $\lambda$-constacyclic code over R is a $\lambda^{-1}-$constacyclic code (\cite{5, 10}).
Let $f(x)$ be the polynomial $f(x)= a_0+a_1x+...+a_r x^r \in R[x]$, and i be the smallest integer such that $a_i \neq 0$. The reciprocal polynomial of f  denoted by $f^*$ is defined as $f^*(x)= x^{r+i}f(x^{-1})=a_r x^{i}+a_{r-1}x^{i+1}+...+a_i x^{r}$.
The following result characterizes the dual of a $\lambda-$ constacyclic code over R.
\begin{theo}
Let $ \C= e_1 \C_1 \oplus e_2\C_2 \oplus e_3 \C_3 \oplus e_4 \C_4$ be a $\lambda-$ constacyclic code of length n over R, such that $\C=<e_1g_1(x), e_2 g_2(x), e_3 g_3(x), e_4 g_4(x)>$ and $\C^{\bot}$ its dual. Let $h_i(x) \in \F_q[x]$ such that $g_i(x)h_i(x)= x^n - \lambda_i.$ Then  $ \C^{\bot}= e_1 \C_1^{\bot} \oplus e_2\C_2^{\bot} \oplus e_3 \C_3^{\bot} \oplus e_4 \C_4^{\bot},$ where $\C_i^{\bot}$ is the dual of the $\lambda_i-$ constacyclic code over $\F_q.$
Furthermore\\ $\C^{\bot}=<e_1 h_1^*(x)+ e_2 h_2^*(x)+ e_3 h_3^*(x) + e_4 h_4^*(x)>.$
\end{theo}
\pr
Let $ s_i \in \C_i^{\bot}, 1 \leq i \leq 4$ and $r=\sum_{i=1}^4 e_i r_i \in \C$ with $r_i \in \C_i, 1 \leq i \leq 4.$ We have that:
$r.(\sum_{i=1}^4 e_i s_i)=(\sum_{i=1}^4 e_i r_i)(\sum_{i=1}^4 e_i s_i)=\sum_{i=1}^4 e_i r_i s_i=0.$ This implies that $e_1 \C_1^{\bot} \oplus e_2\C_2^{\bot} \oplus e_3 \C_3^{\bot} \oplus e_4^{\bot} \C_4 \subseteq \C^{\bot}.$
Note that $|e_1 \C_1^{\bot} \oplus e_2\C_2^{\bot} \oplus e_3 \C_3^{\bot} \oplus e_4 \C_4^{\bot}|= |\C_1^{\bot}||\C_2^{\bot}||\C_3^{\bot}||\C_4^{\bot}|.$ Since R is a finite frobenius ring, then $|\C||\C^{\bot}|=|R|^n$. So :
$$|\C^{\bot}|=\frac{|R|^n}{|\C|}=\frac{q^{4n}}{q^{4n-\sum_{i=1}^4 \deg g_i(x)}}= q^{\sum_{i=1}^4 \deg g_i(x)}$$
$$ = |e_1 \C_1^{\bot} \oplus e_2\C_2^{\bot} \oplus e_3 \C_3^{\bot} \oplus e_4 \C_4^{\bot}|.$$
Hence $\C^{\bot}=e_1 \C_1^{\bot} \oplus e_2\C_2^{\bot} \oplus e_3 \C_3^{\bot} \oplus e_4 \C_4^{\bot}.$

Let $h_i(x) \in \F_q[x]$ such that $g_i(x)h_i(x)= x^n - \lambda_i.$ Since $\C_i$ is a $\lambda_i-$ constacyclic code of length n over $\F_q$, with generator polynomial $g_i(x),$ then $\C_i^{\bot}$ is a $\lambda_i^{-1}-$ constacyclic code with generator polynomial $h_i^*(x)$ that we can suppose monic. From theorem \ref{i3} and theorem \ref{i4}, we conclude that $\C^{\bot}=<e_1 h_1^*(x) + e_2 h_2^*(x)+ e_3 h_3^*(x)+ e_4 h_4^*(x)>.$ \hfill $\square$\\
It's well-known that a $\lambda-$ constacylic code over a finite field can be self-dual if and only if $\lambda^{2}=1.$ So the only self-dual constacyclic codes over finite fields are cyclic and negacyclic codes. Then $\C$ is self-dual code over R if and only if each $\C_i$ is a self-dual cyclic or negacyclic code over $\F_q$ if and only if  $g_i(x)$ and $h_i^*(x)$ are associate in $\F_q [X],$  where $h_i(x)g_i(x)=x^n-\lambda_i$ in $\F_q[x],$ with $\lambda_i=\pm 1.$\\

\section{Gray map with applications}
We define a gray map $\phi_1: R \longrightarrow \F_q^{4}$ by $$\phi_1 (a+bu+cv+duv)= (d, c+d, b+d, a+b+c+d).$$ This map can be extended to $R^n$ in a natural way:
$$
\begin{array}{c c c c }
\phi :& R^n &\longrightarrow & \F_q^{4n} \\
         &(r_0, r_1,...,r_{n-1}) &\longmapsto & (\phi_1(r_1),\phi_1(r_2),...,\phi_1(r_{n-1})).
\end{array}$$

For any element $a+bu+cv+duv \in R,$ we define the Lee weight, denoted by $W_L,$ as $W_L(a+ub+cv+duv)=W_H(d,c+d,b+d,a+b+c+d),$ where $W_H$ denotes the ordinary Hamming weight for q-ary codes. The Lee weight of a codeword $r=(r_1,r_2,...,r_{n-1}) \in R^n$ is defined as $W_L(r)= \sum_{i=0}^{n-1}W_L(r_i)$ and for $r,r^{'} \in R^n,$ the Lee distance is defined as $d_L(r,r^{'})=W_L(r-r^{'}).$ The minimum Lee distance is defined as $min\{d_L(r,r^{'}) | r,r^{'} \in \C, r \neq r^{'}\}.$ We denote the Hamming distance of a q-ary code $\C$ by $d_H(\C).$
The following two results are obvious.
\begin{pro}
The Gray map $\phi$ is a $\F_q-$ linear distance-preserving map from ($R^n,$ Lee distance) to ($\F_q^{4n}$, Hamming distance).
\end{pro}
\begin{lem}
Let $ \C= e_1 \C_1 \oplus e_2\C_2 \oplus e_3 \C_3 \oplus e_4 \C_4$ be a linear code of length n over R, size $q^k$ and minimum Lee distance $d_L,$ then $\phi(\C)$ is a $[4n, k, d_L]-$ linear code over $\F_q.$
\end{lem}
\begin{theo}
Let $ \C= e_1 \C_1 \oplus e_2\C_2 \oplus e_3 \C_3 \oplus e_4 \C_4$ be a linear code of length n over R. Then
$$d_L(\C)= min\{d_H(\C_1, 4 d_H(\C_2), 2 d_H(\C_3), 2 d_H(\C_4)\}.$$
\end{theo}
\pr
The result is obvious because for any codeword $r \in \C, $ we have:
$$\phi(r)=\phi(e_1r_1 + e_2r_2 + e_3r_3 + e_4r_4)= (r_1+r_2-r_3-r_4,r_2-r_3,r_2-r_4,r_2),$$
where $r_1 \in \C_1, r_2 \in \C_2, r_3 \in \C_3, r_4 \in \C_4.$ 
\hfill $\square$\\
Let $\sigma$ be the cyclic shift on $R^n$ defined by $\sigma(r_0,r_1,...,r_{n-1})=(r_{n-1},...,r_{n-2})$ and $n=n^{'}l.$ A linear code which is invariant under $\sigma^l$ is called a l-quasi-cyclic code of length n.
\begin{theo}
A linear code $ \C$ of length n over R is a cyclic code if and only if $\phi(\C)$ is a $4-$ quasi-cyclic code of length $4n$ over $\F_q.$   
\end{theo}
\pr
Let $r=(r_0^{'},r_1^{'},...,r_{n-1}^{'}) \in R^n,$ where $r_i^{'}=a_i+ b_i u +c_i v+d_i uv$ with $a_i, b_i, c_i, d_i \in \F_q, 0 \leq i \leq n-1.$ A simple calculation shows that :
$$ \phi(\sigma(r))=\sigma^4(\phi(r)).$$ Then if $\C$ is a cyclic code of length n over R, we have:
$$\sigma^4(\phi(\C))=\phi(\sigma(\C))=\phi(\C).$$ This implies that $\phi(\C)$ is $4-$ quasi cyclic code of length $4n$ over $\F_q.$\\
The other case is obvious because $\phi$ is an injection.
\hfill $\square$\\

From \cite{9,8}, we know that there exists self-dual cyclic codes of length n  over a finite field $\F_q$ if and only if n is even and q is a power of 2.

\begin{theo}
Let $ \C= e_1 \C_1 \oplus e_2\C_2 \oplus e_3 \C_3 \oplus e_4 \C_4$ be a self-dual cyclic code over $R_2=\F_{2^m}+u\F_{2^m}+v\F_{2^m}+uv\F_{2^m}.$
Then $\phi(\C)$ is a self-dual $4-$ quasi-cyclic code over $\F_{2^m}.$
\end{theo}
\pr
If $r_1,r_2 \in \C,$ then they can be written as follows:\\
$r_1=e_1a_1g_1+e_2a_2g_2+ e_3a_3g_3+e_4a_4g_4; r_2=e_1b_1g_1+e_2b_2g_2+ e_3b_3g_3+e_4b_4g_4;$
where $a_1,a_2,a_3,a_4,b_1,b_2,b_3,b_4 \in R_2$ and $g_1, g_2, g_3, g_4$ are generator polynomials of $\C_1,\C_2,\C_3,\C_4,$ respectively.
If $\C$ is self-dual code over $R_2$, then each $\C_i$ is self-dual cyclic code over $\F_{2^m}$, so $g_i^2=0$ in $\F_{2^m}[x]/<x^n-1>$ and also in $R_2[x]/<x^n-1>.$ Using this fact and because $R_2$ has characteristic 2, we easily check that:$ \phi(r_1).\phi(r_2)=0;$ where\\
$\phi(r_1)=(a_1g_1+a_2g_2-a_3g_3-a_4g_4,a_2g_2-a_3g_3,a_2g_2-a_4g_4,a_2g_2)$ and \\ $\phi(r_2)=(b_1g_1+b_2g_2-b_3g_3-b_4g_4,b_2g_2-b_3g_3,b_2g_2-b_4g_4,b_2g_2).$
\hfill $\square$\\
Now, we give some examples of self-dual cyclic codes over $R_2$ and their Gray images to illustrate the above results.
\begin{exa}
A self-dual cyclic code of length $14$ over $R_2=\F_2+u\F_2+v\F_2+uv\F_2.$\\
The factorisation of $x^{14}+1$ over $\F_2$ is given by:
$$x^{14}+1=(x+1)^2(x^3+x+1)^2(x^3+x^2+1)^2.$$
Let $\C$ be the cyclic code over $R_2$ generated by: $g(x)=e_1g_1(x)+e_2g_2(x)+e_3g_3(x)+e_4g_4(x),$
where $g_1(x)=g_3(x)=x^7+x^6+x^3+x^2+x+1, g_2(x)=x^7+1$ and $g_4(x)=x^7+x^6+x^5+x^4+x+1.$
The codes $\C_1=\C_3=g_1(x)\F_2[x]$ are $[14,7,4]$ self-dual cyclic, $\C_2= g_2(x)\F_2[x]$ is $[14,7,2]$self-dual cyclic and $\C_4=g_4(x)\F_2[x]$ is $[14,7,4]$ self-dual cyclic. Then $\phi(\C)$ is a $[56,28,4]$ self-dual $4-$ quasi-cyclic code over $\F_2.$
\end{exa}
\begin{exa}
A self-dual cyclic code of length $6$ over $R_2=\F_4+u\F_4+v\F_4+uv\F_4.$\\
The factorisation of $x^6+1$ over $\F_4=\F_2[\alpha]$ is given by:
$$x^6+1=(x+1)^2(x+\alpha)^2(x+\alpha^2)^2.$$
Let $\C$ be the cyclic code over $R_2$ generated by: $g(x)=e_1g_1(x)+e_2g_2(x)+e_3g_3(x)+e_4g_4(x),$
where $g_1(x)=g_2(x)=\alpha^2+\alpha^2 x+x^2+x^3, g_3(x)=\alpha+ \alpha x+ x^2+x^3$ and $g_4(x)=x^3+1.$
The codes $\C_1=\C_2=g_1(x)\F_4[x]$ are $[6,3,3]$ self-dual cyclic, $\C_3= g_3(x)\F_4[x]$ is $[6,3,3]$self-dual cyclic and $\C_4=g_4(x)\F_4[x]$ is $[6,3,2]$ self-dual cyclic. Then $\phi(\C)$ is a $[24,12,3]$ self-dual $4-$ quasi-cyclic code over $\F_4.$
\end{exa}

\section{Conclusion}
In this paper, the generator polynomials of constacyclic codes over $\F_q+u\F_q+v\F_q+uv\F_q,$ and their duals are characterized, with help of some decomposition of the ring. We have also given a necessary and sufficient condition on the existence of self-dual constacyclic codes. We have shown that the Gray image of a self-dual cyclic code of length n over $\F_{2^m}+u\F_{2^m}+v\F_{2^m}+uv\F_{2^m}$ is a self-dual $4-$quasi cyclic code of length $4n$ over $\F_{2^m}.$
%

\end{document}